\documentclass{article}
\usepackage{spconf,amsmath,graphicx,multirow,bm}
\usepackage{tikz,pgfplots}
\usepackage{pgfplotstable}
\pgfplotsset{filter discard warning=false}

\title{CONFIDENCE ESTIMATION AND DELETION PREDICTION USING\\BIDIRECTIONAL RECURRENT NEURAL NETWORKS}
\name{A.~Ragni, Q.~Li, M.~J.~F. Gales, Y.~Wang\thanks{
This research is based upon work supported in part by the ALTA Institute, 
Cambridge University and the Office of the
Director of National Intelligence (ODNI), Intelligence Advanced Research
Projects Activity (IARPA), via Air Force Research Laboratory (AFRL) 
contract \# FA8650-17-C-9117. The views and
conclusions contained herein are those of the authors and should not be
interpreted as necessarily representing the official policies, either
expressed or implied, of ODNI, IARPA, AFRL or the U.S. Government. The U.S.
Government is authorized to reproduce and distribute reprints for
governmental purposes notwithstanding any copyright annotation therein.
}}
\address{Department of Engineering, University of Cambridge\\
Trumpington Street, Cambridge CB2 1PZ, UK\\
{\tt \{ar527,ql264,mjfg,yw396\}@eng.cam.ac.uk}}

\begin{document}
\ninept
\maketitle
\begin{abstract}
The standard approach to assess reliability of automatic speech
transcriptions is through the use of confidence scores. 
If accurate, these scores provide a flexible mechanism to flag 
transcription errors for upstream and downstream applications. 
One challenging type of errors that recognisers make are 
deletions. These errors are not accounted for by 
the standard confidence estimation schemes and are hard 
to rectify in the upstream and downstream processing. High 
deletion rates are prominent in limited resource 
and highly mismatched training/testing conditions studied 
under IARPA Babel and Material programs. 
This paper looks at the use of bidirectional recurrent neural
networks to yield confidence estimates in predicted 
as well as deleted words. Several simple schemes are examined 
for combination. To assess usefulness of this 
approach, the combined confidence score is examined for 
untranscribed data selection that favours transcriptions with 
lower deletion errors. Experiments are conducted using IARPA 
Babel/Material program languages.
\end{abstract}
\section{Introduction}
\label{sec:intro}
Recent years have seen an increase in demand for speech enabled solutions. These range from speech transcription to personal assistants \cite{Li2017aS,Ram2018aS} designed to handle ever increasing in complexity human-machine interactions \cite{Young2010aS}. The perceived usefulness of these applications depends on the quality of the underlying automatic speech recogniser which for some tasks has been reported to approach the agreement level of human annotators \cite{Saon2016aS,Xiong2017aS}. Even in such favourable conditions a measure of reliability in hypothesised words can prove useful for flagging up words that may need to be handled differently. For more challenging scenarios such measure can prove fundamental for achieving high performance \cite{Alumae2017aS} and yielding reliable feedback \cite{Knill2018aS}. The development cycle of speech recognisers may itself benefit from the measure of reliability. Rather than expending resources on annotating data that can already be reliably transcribed a more challenging set can be automatically identified in a live data and incorporated into training these recognisers. 

Confidence scores \cite{Jiang2005aL} have been traditionally used as the measure of reliability of automatic speech transcriptions in both downstream as well as upstream tasks. In the simplest form these scores are posterior probabilities associated with words in hypothesised transcriptions. These probabilities however are believed to be over estimated \cite{Evermann2000aL} which could hamper their utility as the confidence measure. A range of approaches have been examined to improve accuracy of confidence estimates. This includes simple piece-wise linear mappings in the form of decision trees \cite{Evermann2000aL}, generalised linear models \cite{Gillick1997aS}, highly non-linear feed-forward neural networks \cite{Weintraub1997aS} and sequence models, such as conditional random fields (CRF) \cite {Seigel2011aS, Sutton2010aL} and recurrent neural networks \cite{Kalgaonkar2015aS}. All these approaches reported gains over posterior based confidence scores using a range of evaluation metrics. This paper examines a neural network model that takes into account information about not only the current word and its history but also about all the following words. The bidirectional nature of this model is expected to be well suited for the task of confidence estimation given the complete knowledge of the entire hypothesised sequence. This work thus combines the benefits of neural networks for feature extraction and context modelling with the whole sequence context of CRFs in a single model.\footnote{This part of the work was completed prior to the publication of an independent study on confidence prediction using BiRNNs \cite{DelAgua2018aS}.}

The standard formulation of confidence scores is focused on assigning estimates to hypothesised words only. This leaves any deleted word out of consideration. Deletions, however, are a very challenging type of error to recover from in both downstream and upstream tasks. Moreover, drifts in deployment domains either planned \cite{Ragni2018aS} or unplanned \cite{Strope2011aS} typically lead to large increases in deletion errors. Though schemes such as frame weighting \cite{Chan2004aS, Ma2006aS} could be used to alleviate the impact of deletions to a certain extent, an explicit deletion prediction is expected to be more appropriate. The previous work on deletion prediction \cite{Seigel2014aS} has looked at predicting whether a deletion occurs between current and next word using CRFs. This paper extends that work by predicting deletions by means of a bidirectional recurrent neural network (BiRNN) model \cite{Schuster1997aS}. For these deletion predictions to be useful a mechanism is needed to incorporate them in upstream and downstream tasks. This paper examines utterance selection for unsupervised training as the target task and shows how simple combinations of confidence and deletion predictions can yield selected data with different deletion error rate characteristics. Experiments with IARPA Babel/Material program languages show promise of this approach for confidence estimation and deletion prediction.

The rest of this paper is organised as follows. Section~\ref{sec:confidence} discusses confidences and evaluation metrics. The following section~\ref{sec:deletion} examines interactions between these scores and deletions. Section~\ref{sec:birnn} describes the proposed BiRNN architecture for confidence and deletion prediction. Experimental results are presented in Section~\ref{sec:experiments}. Finally, conclusions drawn from this work are given in Section~\ref{sec:conclusions}. 

\section{Confidence scores}
\label{sec:confidence}
Quantifying uncertainty in predictions is a challenge that modern speech recognisers are faced with due to their ubiquitous use in an ever growing number of applications. Confidence scores have been a traditional measure of certainty of a speech recogniser in its prediction \cite{Jiang2005aL}. These scores are typically derived for each hypothesised word. The simplest form of a confidence score is a posterior probability of a hypothesised word that can be estimated over a lattice of possible transcriptions generated by the speech recogniser \cite{Kemp1997aS, Wessel1998aS}. This simple approach has been found \cite{Evermann2000aL} to significantly over-estimate these scores. The reason most often stated is a limited size of lattices over which these posterior probabilities are normalised. In order to adjust confidences so they better reflect the true values a number of approaches have been examined. These can be divided into simple mappings that make use of only original scores and more complex approaches that make use of additional information. The most prominent example of the former are decision trees \cite{Evermann2000aL}. This simple approach estimates a piece-wise linear mapping given some held-out data. The latter group, including the work reported in this paper, make use of more complex approaches, such as CRFs \cite{Seigel2011aS}, feed-forward \cite{Weintraub1997aS} and recurrent neural networks \cite{Kalgaonkar2015aS} to estimate confidence scores from manually designed features. These features may include various statistics extracted from audio, acoustic models, language models and lattices \cite{Schaaf1997aS}. 

In the ideal, limiting, case, confidence scores of correctly hypothesised words are one and zero otherwise. In order to measure how far any given set of confidence scores is from the ideal scenario a number of metrics could be used. One popular metric is a normalised cross-entropy (NCE) \cite{Siu1997aS}. This metric measures a relative change in the cross-entropy caused when an empirical estimate of ASR correctness is replaced by hypothesised confidences. The estimate of ASR correctness can be computed from reference confidences ${\bf c}^{*}=\{c_{1}^{*},\ldots,c_{T}^{*}\}$ by
\begin{equation}
P(C=1) = \frac{1}{T} \sum_{t=1}^{T} c_{t}^{*}
\end{equation}
where $P(C=1)$ is the probability of ASR system being correct and $P(C=0)=1-P(C=1)$ is for the opposite event, $T$ is the number of hypothesised words. The average binary cross-entropy between this estimate and reference confidences is given by
\begin{equation}
\overline{H}({\bf c}^{*}) = -\frac{1}{T}\sum_{t=1}^{T} c_{t}^{*}\log(P(C=1)) + (1-c_{t}^{*})\log(P(C=0))  
\end{equation}
On the other hand, the average binary cross-entropy between hypothesised ${\bf c}=\{c_{1},\ldots,c_{T}\}$ and reference confidences is given by
\begin{equation}
H({\bf c}|{\bf c}^{*}) = - \frac{1}{T}\sum_{t=1}^{T} c_{t}^{*}\log(c_{t}) + (1 - c_{t}^{*})\log(1-c_{t})
\end{equation}
If the hypothesised confidences ${\bf c}$ are systematically better than the estimate of ASR system being correct, $P(C=1)$, the relative change in cross-entropy or NCE given by
\begin{equation}
\text{NCE}({\bf c},{\bf c}^{*}) = \frac{\overline{H}({\bf c}^{*}) - H({\bf c}|{\bf c}^{*})}{\overline{H}({\bf c}^{*})}
\end{equation}
is positive. In the opposite case it is negative. The maximum value of NCE is 1 which corresponds to the case where hypothesised confidences match reference confidences exactly. 

Though popular NCE may not be the most optimal metric to assess confidence scores. For these scores to be the perfect correct/incorrect predictor, it is not necessary to match reference confidences exactly. It is however sufficient to yield correct rank ordering such that confidences of all incorrectly hypothesised words are below some threshold and those of all correctly hypothesised words are above. This suggests that an area under the curve (AUC) type metric may be more suitable. When choosing which curve to use it is important to know the balance of positive and negative class examples. For balanced data it is common to use a receiver operating curve (ROC) of false positive (FP) rate (FPR) against true positive (TP) rate (TPR) for a range of threshold $\theta$ values
\begin{equation}
\text{TPR}(\theta) = \frac{\text{TP}(\theta)}{\text{TP}(\theta)+\text{FN}(\theta)},\qquad \text{FPR}(\theta)=\frac{\text{FP}(\theta)}{\text{FP}(\theta)+\text{TN}(\theta)}
\end{equation}
where $\text{TN}(\theta)$ and $\text{FN}(\theta)$ are true negatives and false negatives. The AUC of a random guess under the ROC curve is 0.5. Values larger than that signal better than random performance. For imbalanced data it is more appropriate to use a precision-recall curve \cite{Davis2006bS}
\begin{equation}
\text{Precision}({\theta})=\frac{\text{TP}(\theta)}{\text{TP}(\theta)+\text{FP}(\theta)},\quad \text{Recall}(\theta)=\frac{\text{TP}(\theta)}{\text{TP}(\theta)+\text{FN}(\theta)}
\end{equation}
The AUC associated with a random guess under the precision recall curve depends on the ratio between positive $P$ and negative $N$ examples and is given by $\frac{P}{P+N}$. For a perfectly balanced data set this yields $0.5$ and less and larger than $0.5$ for data sets dominated by negative and positive examples respectively. Speech recognisers described in this work operate in the region where there are more correctly than incorrectly hypothesised words. However, as the difference is not large the ROC curve will be used in this work. 

Confidence scores are used not only in downstream applications but also in upstream tasks. For instance in semi-supervised training confidence scores are routinely used for selecting high-confidence hypotheses for training acoustic models \cite{Ma2006aS}. This requires extending the definition of confidence scores to an utterance (segment) and audio recording level. Given a sequence of confidence scores ${\bf c}_{r} = \{c_{r,1},\ldots,c_{r,T_r}\}$ associated with recording $r$, a common solution is to compute frame-weighted confidence score \cite{Chan2004aS, Ma2006aS}
\begin{equation}
c_{r} = \frac{\sum_{t=1}^{T_r} \lambda_{r,t} c_{r,t}}{\sum_{t=1}^{T_r} \lambda_{r,t}}
\label{eq:frame_weighting}
\end{equation}
where $T_r$ is the number of hypothesised words and $\lambda_{r,t}$ is the number of frames associated with hypothesised word $w_{r,t}$. This enables the complete dataset to be rank ordered and manipulated to yield subsets of the data with suitable confidence characteristics. The frame-weighting applied in equation~\eqref{eq:frame_weighting} ensures that long, high confidence, regions do not get penalised unfairly by short, low confidence, regions. The fundamental problem with using equation~\eqref{eq:frame_weighting} for data selection is that it does not take into account deletion errors and may select transcriptions embedding large numbers of deletions. 

\section{Deletion errors}
\label{sec:deletion}
Errors made by a speech recogniser are traditionally partitioned into 3 types: substitution, insertion and deletion. The former two have a direct realisation in the form of hypothesised words whereas deletions do not. The precise nature of partitioning into these error types is a complex issue. First, the errors are computed by a Levenshtein alignment that weighs different types unequally \cite{htkbook2015aL}. This balance is further affected non-uniformly by the task at hand, the forms of acoustic and language models used, decoding approaches. Any mismatch between training and testing conditions typically would lead to an additional imbalance. Although some forms of imbalance may not necessary be harmful, high levels of deletion errors do not interact well with a range of techniques such as discriminative adaptation/training, semi-supervised training, forms of minimum Bayes' risk decoding. Therefore, keeping deletion error under control is an important attribute of building speech recognisers. 

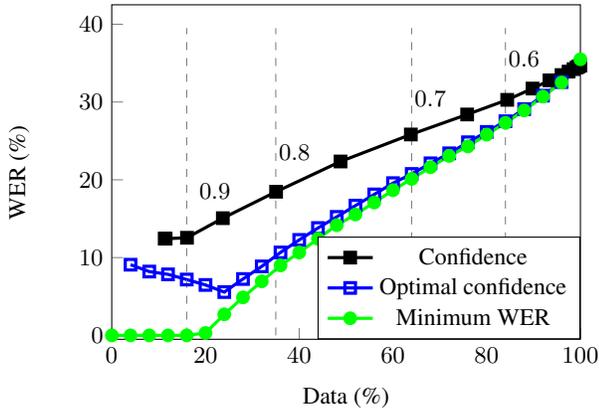
\begin{figure}[!htbp]
\centering
\begin{tikzpicture}
\begin{axis}[width=0.35\textwidth,height=0.25\textwidth,scale only axis,xlabel=Data (\%),ylabel=WER (\%),xmin=0,xmax=100,ymin=-0.5,ymax=42.5,legend style={at={(1.0,0.0)},anchor=south east},xtick pos=left,ytick pos=left]
\addplot[very thick,color=black,mark=square*] table [x expr=100.0*\thisrow{D}/52293.55, y=WER]{fig_used/wer/swahili_evalpart1_manseg_map.dat};
\addlegendentry{Confidence}
\addplot[very thick,color=blue,mark=square] table [x=D, y=WER]{fig_used/swahili_dev_dur_sub_ins.dat};
\addlegendentry{Optimal confidence}
\addplot[very thick,color=green,mark=*] table [x=D, y=WER]{fig_used/swahili_dev_dur_wer.dat};
\addlegendentry{Minimum WER}
\draw[color=gray,dashed] (16,5) -- (16,400);
\draw (22,190) node {$0.9$};
\draw[color=gray,dashed] (35,5) -- (35,400);
\draw (39,240) node {$0.8$};
\draw[color=gray,dashed] (64,5) -- (64,400);
\draw (68,310) node {$0.7$};
\draw[color=gray,dashed] (84,5) -- (84,400);
\draw (88,360) node {$0.6$};
\end{axis}
\end{tikzpicture}
\caption{Example impact of data selection approaches on word error rate (WER) in selected transcriptions}
\label{fig:example_confidence}
\end{figure}

\begin{figure}[!htbp]
\centering
\begin{tikzpicture}
\begin{axis}[width=0.35\textwidth,height=0.25\textwidth,scale only axis,xlabel=Data (\%),ylabel=Relative Error Reduction (\%),xmin=0,xmax=100,ymin=0.0,ymax=100.0,legend style={at={(1.0,0.51)},anchor=south east},xtick pos=left,ytick pos=left]
\addplot[very thick,color=black,mark=square*] table [x expr=100.0*\thisrow{Dur}/52293.55, y=Sub]{fig_used/swahili_evalpart1_manseg_relative.dat};
\addlegendentry{Substitution}
\addplot[very thick,color=blue,mark=square] table [x expr=100.0*\thisrow{Dur}/52293.55, y=Del]{fig_used/swahili_evalpart1_manseg_relative.dat};
\addlegendentry{Deletion}
\addplot[very thick,color=green,mark=*] table [x expr=100.0*\thisrow{Dur}/52293.55, y=Ins]{fig_used/swahili_evalpart1_manseg_relative.dat};
\addlegendentry{Insertion}
\addplot[very thick,color=red,mark=o] table [x expr=100.0*\thisrow{Dur}/52293.55, y=WER]{fig_used/swahili_evalpart1_manseg_relative.dat};
\addlegendentry{Total}
\addplot[thick,color=gray,mark=-,dashed] table [x=X,y=Y]{fig_used/dummy.dat};
\addlegendentry{Confidence}
\draw[color=gray,dashed] (16,5) -- (16,95);
\draw (20,85) node {$0.9$};
\draw[color=gray,dashed] (35,5) -- (35,95);
\draw (39,60) node {$0.8$};
\draw[color=gray,dashed] (64,5) -- (64,95);
\draw (68,35) node {$0.7$};
\draw[color=gray,dashed] (84,5) -- (84,95);
\draw (88,20) node {$0.6$};
\end{axis}
\end{tikzpicture}
\caption{Example of relative reduction in substitution, deletion, insertion and total error in confidence-based data selection}
\label{fig:example_relative}
\end{figure}
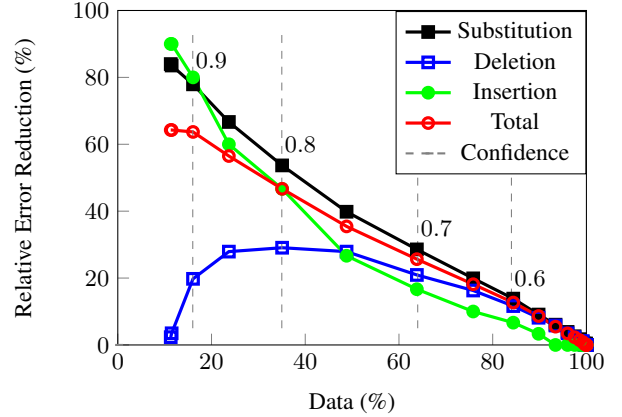

The standard confidence estimation and confidence-based approaches do not explicitly address the presence of deletions \cite{Seigel2014aS}. This may not necessary lead to a problem however. Figure~\ref{fig:example_confidence} shows an example of the impact that the average frame-weighted confidence-based data selection has on word error rate (WER). The underlying is a narrow-band conversational telephone speech (CTS) recogniser for a limited resource language.  
As can be seen from Figure~\ref{fig:example_confidence}, the WER of the full data set is above 30\%. The top, filled squares, line shows that as the amount of selected data is decreased, by imposing ever increasing minimum average frame-weighted confidence constraints, the WER decreases approaching 10\%. The middle, hollow squares, oracle line shows what would have happened if correct confidences were available. The bottom, filled circles, oracle line shows the impact of also using the knowledge of deletion errors. These lines show that large improvements are possible if better confidence scores were used and unless limited quantities of data are selected the impact of deletion errors can be small. The latter observation however does not take into account the presence of correlation between substitutions, insertions and deletion errors that can be exploited if deletions were taken into account by the data selection scheme. Figure~\ref{fig:example_relative} provides an insight into the individual error distributions during the selection process. 
As the amount of confidence required to be selected increases the relative reduction in all error types increases up to 0.8 average frame-weighted confidence level where the speech recogniser starts to delete progressively more words. Thus, the transcriptions that the speech recogniser is most confident about may have worst deletion characteristics. Table~\ref{tab:cross_domain} shows what happens when this narrow-band speech recogniser is used to bootstrap a wide-band speech recogniser on some untranscribed news broadcast data. 
\begin{table}[!htbp]
\centering
\begin{tabular}{ccc||cccc}
\hline
\multirow{2}*{Type} & \multicolumn{2}{c||}{Band} & \multicolumn{4}{c}{Error (\%)}\\
 & Model & Data & Sub & Del & Ins & Tot\\
\hline\hline
\multirow{2}*{Sup} & Narrow & Narrow & 24.1 & 8.3 & 3.1 & 35.5\\
 & Narrow & Wide & 23.4 & 19.0 & 1.3 & 43.7\\
Unsup & Wide & Wide & 15.6 & 25.5 & 0.9 & 42.0\\
\hline
\end{tabular}
\caption{Impact of band mismatch on unsupervised training}
\label{tab:cross_domain}
\end{table}
The first line shows the narrow-band recogniser error characteristics on a set of narrow-band development data. The middle line shows its performance on a down-sampled set of wide-band data containing news and topical broadcasts. As can be seen this mismatch has largely affected deletions. The last line shows that automatic transcriptions produced in such mismatched configuration, when used to train a wide-band recogniser, lead to an additional increase in deletions. Thus schemes that can lower deletions in selected automatic transcriptions would be of interest in such mismatched configurations.

If confidence estimates are supplemented with deletion estimates the standard confidence-based selection scheme needs to be altered. A range of schemes can be devised that weigh confidences and deletions in a number of different ways. One approach would be to discount confidences in equation~\eqref{eq:frame_weighting} by scaled deletion estimates
\begin{eqnarray}
\hat{c}_{t} &=& c_{t} - \theta_d d_{t}\label{eq:discount1}\\
\hat{c}_{1} &=& c_{1} - \theta_d d_{1} - \theta_{s}s\label{eq:discount2}
\end{eqnarray}
This scheme has two free parameters and degenerates to the standard confidence-based selection scheme when $\theta_d$ and $\theta_s$ are zero. An alternative approach would be to threshold both estimates to yield estimates of correct hypotheses
\begin{equation}
\text{Cor}({\bf c};\theta_{c}) = \sum_{t=1}^{T} \delta(c_{t}\geq\theta_{c})
\label{eq:threshold_cor}
\end{equation}
substitution and insertion errors
\begin{equation}
\text{Inc}({\bf c};\theta_{c}) = \sum_{t=1}^{T} \delta(c_{t}<\theta_{c})
\label{eq:threshold_inc}
\end{equation}
as well as deletions
\begin{equation}
\text{Del}({\bf d};\theta_{s},\theta_{d}) = \delta(s\geq\theta_{s}) + \sum_{t=1}^{T} \delta(d_{t}\geq\theta_{d})
\label{eq:threshold_del}
\end{equation}
where $\theta_{c}$ and $\theta_{s}, \theta_d$ are thresholds for correct hypotheses and deletions, $d_{t}$ is an estimate of deleting a word following the hypothesis at time $t$ and $s$ is an estimate of deleting a word prior to the first hypothesised word. The confidence and deletion based data selection can be performed by rank ordering utterances/recordings according to an estimate of WER given by
\begin{equation}
\text{WER}({\bf c},{\bf d};{\bm\theta}) = \frac{\text{Inc}({\bf c};\theta_c) + \text{Del}({\bf d};\theta_s,\theta_d)}{\theta_p\cdot\text{Inc}({\bf c};\theta_c) + \text{Cor}({\bf c};\theta_c)}
\label{eq:threshold_wer}
\end{equation}
where $\theta_p$ is a penalty term to reduce over-estimation of reference words in the denominator since $\text{Inc}({\bf c};\theta_c)$ counts both substitution and insertion errors. Omitting $\text{Inc}({\bf c};\theta_c)$ term in the numerator would enable to perform data selection based on the estimate of deletion error alone, however, some safe guarding mechanism would be required to ensure that the overall WER is low. The thresholds ${\bm\theta}$ can be optimised by minimising mean squared error between the true and predicted WER on some held-out data using, for instance, a grid search or a constrained derivative free method \cite{Ragni2018aS}. 

\section{Bidirectional recurrent neural networks}
\label{sec:birnn}
\begin{figure}[htbp]
\centering
\begin{minipage}{0.4\columnwidth}
\includegraphics[width=2.5cm]{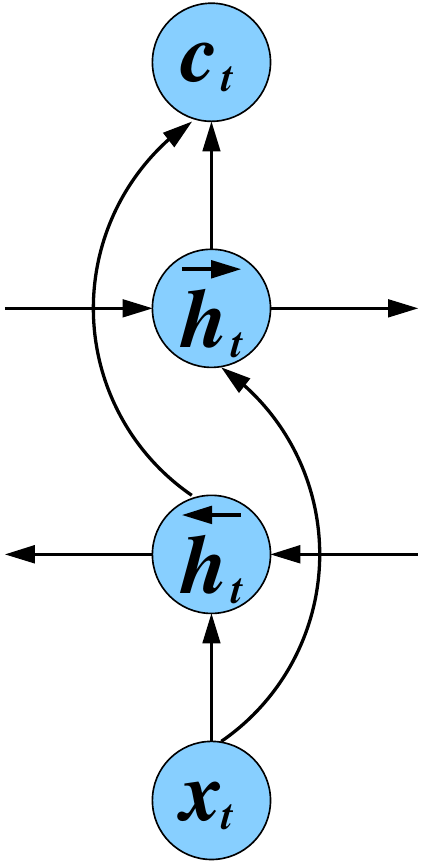}
\centerline{(a) confidence prediction}
\end{minipage}
\begin{minipage}{0.4\columnwidth}
\includegraphics[width=3.0cm]{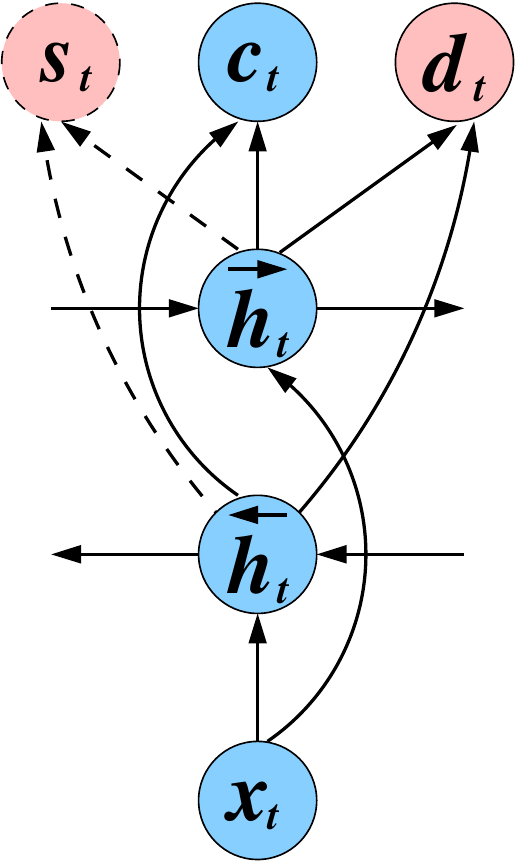}
\centerline{(b) deletion prediction}
\end{minipage}
\caption{Bidirectional recurrent neural networks}
\label{fig:birnn}
\end{figure}

In this work bidirectional recurrent neural networks (BiRNN) \cite{Schuster1997aS} are used for estimating confidences and deletions. Figure~\ref{fig:birnn}~(a) shows the simplest form of BiRNNs that predicts confidence at time $t$ by means of two recurrent units modelling past $\overrightarrow{\bf h}_{t}$ and future $\overleftarrow{\bf h}_{t}$ information.  
Various options exist how the past or history information can be updated at the following time $t+1$. In the simplest case 
\begin{equation}
\overrightarrow{\bf h}_{t+1} = \sigma({\bf W}^{(\overrightarrow{h})}\overrightarrow{\bf h}_{t} + {\bf W}^{(x)}{\bf x}_{t+1})
\end{equation}
where ${\bf x}_{t+1}$ is an input feature vector, ${\bf W}^{(x)}$ is an input matrix, ${\bf W}^{(\overrightarrow{h})}$ is a history matrix and $\sigma$ is a non-linearity such as sigmoid. The future information is typically modelled in the same fashion though this is not a requirement. The past and future information at any time $t$ can be combined into a single context vector 
\begin{equation}
{\bf h}_{t} = \begin{bmatrix}\overrightarrow{\bf h}_{t} & \overleftarrow{\bf h}_{t}\end{bmatrix}^{\sf T}
\end{equation}
Given ${\bf h}_{t}$, the confidence $c_{t}$ can be modelled by
\begin{equation}
c_{t} = \sigma({\bf w}^{(c)^{\sf T}}{\bf h}_{t} + {b}^{(c)})\\
\end{equation}
where ${\bf w}^{(c)}$ is a parameter vector and $b^{(c)}$ is a bias, $\sigma$ is any non-linearity mapping confidences into $[0, 1]$ range.  

One possible way to extend this BiRNN from predicting just confidences to also predicting deletions is shown in Figure~\ref{fig:birnn}~(b). Here, at any time $t$ an additional, deletion, prediction $d_{t}$ is made that a deletion occurred between current and the following word
\begin{equation}
d_{t} = \sigma({\bf w}^{(d)^{\sf T}}{\bf h}_{t} + {b}^{(d)})
\end{equation}
At time $t=1$ an additional prediction is made to yield an estimate of deletion prior to the first hypothesised word
\begin{equation}
s = \sigma({\bf w}^{(s)^{\sf T}}{\bf h}_{1} + {b}^{(s)})
\end{equation}
Not that such definition cannot take into account multiple consecutive deletions. 
Training can be performed by minimising cross-entropies between true and predicted values of confidences and deletions
\begin{equation}
{\mathcal F}({\bm\theta}) = \sum_{r=1}^{R} \left(\sum_{t=1}^{T_{r}} {c}_{r,t}^{*} \log(c_{r,t}) + {d}_{r,t}^{*}\log(d_{r,t})\right)+{s}_{r}^{*}\log(s_{r})\\
\label{eq:ce}
\end{equation}
where the asterisk $*$ denotes true target values. For improved generalisation the objective function above can be regularised by adding a scaled $L_2$ norm of model parameters ${\bm\theta}$. 

For this approach to be successful, the input features should include information about hypothesised word sequence that is correlated with different error types. A large body of work has been produced on hand-crafted features for these general machine learning approaches \cite{Schaaf1997aS, Weintraub1997aS, Ma2007aS, Seigel2014aS}. For this preliminary exploration a simple and compact form was examined for each hypothesised word
\begin{equation}
{\bf x}_t = \begin{bmatrix}
\lambda_t\\
c_t\\
{\bf e}(w_t)\\
{\mathcal O}(w_t|w_{t-1},\ldots)\\
\log(P(w_t|w_{t-1},\ldots))\\
|w_t|\\
\delta_{t-1}\\
\delta_{t+1}
\end{bmatrix}
\label{eq:birnn_features}
\end{equation}
This includes duration $\lambda_t$, confidence $c_t$ as estimated by the decision tree, word $w_t$ embedding ${\bf e}(w_t)$, $n$-gram language model statistics in the form of log-probability $\log(P(w_t|w_{t-1},\ldots))$ and the highest non-backoff $n$-gram order ${\mathcal O}(w_t|w_{t-1},\ldots)$, character length $|w_t|$ and any time gap between the current word and preceding $\delta_{t-1}$ and following $\delta_{t+1}$ word. Though this set is limited it provides a sufficient ground for exploring confidence and deletion prediction. 

\section{Experiments}
\label{sec:experiments}
Experiments were conducted on three languages: Georgian, Tagalog and Swahili. These languages originate from different language families and possess limited resources available for the development of speech recognisers. All three languages were a subject of research under the IARPA BABEL program for developing agile and robust speech recognition technology that can be rapidly applied to any language. The latter two languages have also been included into the new IARPA initiative for machine translation for English retrieval and summarisation of text and speech (MATERIAL). For each language only limited quantities of transcribed narrow-band conversational telephone speech (CTS) data are available. This consists of 40-60 hours for training, 10 hours for development and 10-15 hours for evaluation. The Material program interested in both narrow-band CTS and a range of wide-band domains has also released 10 hours of transcribed wide-band news and topical broadcasts for performers to analyse how well they can handle large domain mismatches. 

All speech recognisers made use of lattice-free maximum mutual information (LF-MMI) interleaved time-delay and long short-term memory neural network (TDNN-LSTM) \cite{Peddinti2017aS, Povey2011aS} acoustic models and $n$-gram language models \cite{Stolcke2002aS}. Unless otherwise stated, all acoustic models were trained on narrow-band CTS data. Input features for acoustic models consisted of 24 dimensional filter bank coefficients, probability of voicing and pitch \cite{Ghahremani2014aS}. All language models were obtained by interpolating multiple source language models on transcriptions of development data. The source language models were built on training data transcriptions and various sources of web data harvested by Babel/Material program performers \cite{Mendels2015aS, Zhang2015aS}. The amount of web data varied from 140 million words for Georgian to 1 billion words for Tagalog and Swahili. Pronunciations for these words were obtained by a grapheme-to-phoneme model \cite{Bisani2008aS}. Table~\ref{tab:asr_cts} shows performance of these recognisers on Babel development ({\tt dev}) and evaluation ({\tt evalpart1}) data. 
\begin{table}[htbp]
\centering
\begin{tabular}{c|c||cccc}
\hline
\multirow{2}*{Language} & \multirow{2}*{Test set} & \multicolumn{4}{c}{Error (\%)}\\
 & & Sub & Del & Ins & Tot\\
\hline\hline
\multirow{2}*{Georgian} & {\footnotesize\tt dev} & 26.7 & 10.3 & 2.9 & 39.9\\
 & {\footnotesize\tt evalpart1} & 22.9 & 8.0 & 2.9 & 33.8\\
\hline\hline
\multirow{2}*{Tagalog} & {\footnotesize\tt dev} & 26.4 & 9.9 & 4.3 & 40.6\\
 & {\footnotesize\tt evalpart1} & 25.6 & 9.4 & 4.2 & 39.2\\
\hline\hline
\multirow{2}*{Swahili} & {\footnotesize\tt dev} & 24.1 & 8.3 & 3.1 & 35.5\\
 & {\footnotesize\tt evalpart1} & 23.4 & 8.2 & 3.4 & 35.0\\
\hline
\end{tabular}
\caption{Speech recognition performance on narrow-band development and evaluation data}
\label{tab:asr_cts}
\end{table}
The total error rate, WER, in these limited resource conditions is high, ranging between 35 and 40\%. For all languages the development data provides a good match to the evaluation data. In particular, the Georgian evaluation data appears to be significantly easier. 

Two approaches for confidence estimation are examined in this work: baseline decision trees (DT) and bidirectional recurrent neural networks (BiRNN) in Figure~\ref{fig:birnn}~(a). The continuous bag of words representation was chosen to train 50-dimensional word embeddings on all available text data using {\tt fastText} library \cite{Bojanowski2016aL}. These embeddings provided the majority of input features in equation~\eqref{eq:birnn_features}. The BiRNN made use of single layer 64-dimensional LSTM units to model the past and future information. The development sets were used to train BiRNNs parameters using {\tt TensorFlow} library \cite{Tensorflow2015aL}. Table~\ref{tab:conf_birnn} contrasts performances of the original confidence scores as given by the recogniser (CN), after decision tree mapping (DT) and after BiRNN estimation.
\begin{table}[htbp]
\centering
\begin{tabular}{c|c||ccc}
\hline
\multirow{2}*{Language} & \multirow{2}*{Metric} & \multicolumn{3}{c}{Confidence Estimation}\\
 & & CN & DT & BiRNN\\
\hline\hline
\multirow{2}*{Georgian} & NCE & -0.376 & 0.256 & 0.282\\
 & AUC & 0.831 & 0.831 & 0.847\\
\hline\hline
\multirow{2}*{Tagalog} & NCE & -0.428 & 0.219 & 0.249\\
 & AUC & 0.810  & 0.810 & 0.825\\
\hline\hline
\multirow{2}*{Swahili} & NCE & -0.297 & 0.372 & 0.392\\
 & AUC & 0.832  & 0.832 & 0.846\\
\hline
\end{tabular}
\caption{Normalised cross-entropy and area under the ROC curve performance on narrow-band evaluation data}
\label{tab:conf_birnn}
\end{table}
As the proportion of positive examples is not high the ROC curve was used to compute areas under the curve (AUC). The CN and DT columns show that decision trees have significantly reduced cross-entropies between true and estimated confidences thus leading to lower normalised cross-entropy (NCE) values. However, the AUC values have not changed due to monotonicity of the learned mappings incapable of changing the rank ordering. The last column shows that BiRNNs examined in this work provide small but consistent gains in the NCE and AUC values. 

For confidence and deletion prediction the topology of BiRNN is adjusted according to Figure~\ref{fig:birnn}~(b). Apart from doubling dimensionality of LSTM units no other changes were made. For consistency with the previous results Table~\ref{tab:del_birnn} shows areas under the ROC curve.
\begin{table}[htbp]
\centering
\setlength\tabcolsep{5pt}
\begin{tabular}{c|c||ccc}
\hline
\multirow{2}*{Predictor} & \multirow{2}*{Prediction} & \multicolumn{3}{c}{Language}\\
 & & Swahili & Tagalog & Georgian\\
\hline\hline
\multirow{1}*{Cor/Inc} & \multirow{1}*{Cor/Inc} 
 & 0.846 & 0.825 & 0.847\\
\hline\hline
\multirow{3}*{+Del}  & \multirow{1}*{Cor/Inc} 
 & 0.846 & 0.826 & 0.847\\
\cline{2-5}
 & \multirow{1}*{Next Del} 
 & 0.742 & 0.740 & 0.726\\
\cline{2-5}
 & \multirow{1}*{Start Del} 
 & 0.746 & 0.646 & 0.644\\
\hline
\end{tabular}
\caption{Area under the ROC curve performance of BiRNNs for correct/incorrect, sentence start and next word deletion prediction on narrow-band evaluation data}
\label{tab:del_birnn}
\end{table}
The first block repeats the results in Table~\ref{tab:conf_birnn}. The first line in the second block shows that the AUC performance remained largely unaffected from the introduction of two additional tasks. The next line shows that predicting whether a deletion occurred before the following word is challenging with AUC values going down by as much as 0.1. The last line shows that, apart from Swahili, predicting that a deletion occurred prior to the first hypothesised word is even more challenging. Both are believed to be the consequence of small number of positive examples and limited features used in this work.

\begin{figure}[!htbp]
\centering
\begin{tikzpicture}
\begin{axis}[width=0.35\textwidth,height=0.25\textwidth,scale only axis,xlabel=Data (\%),ylabel=Deletion (\%),xmin=0,xmax=100,ymin=0,ymax=8.7,legend style={at={(1.0,0.0)},anchor=south east},xtick pos=left,ytick pos=left]
\addplot[very thick,color=black,mark=square*] table [x=D, y=Del]{fig_used/swahili_cn_map_dur_wer_del.dat};
\addlegendentry{Decision tree}
\addplot[very thick,color=blue,mark=square] table [x=Dur, y=Del]{fig_used/tmp2.dat};
\addlegendentry{BiRNN threshold}
\addplot[very thick,color=green,mark=*] table [x=Dur, y=Del]{fig_used/del/swahili_evalpart1_manseg_map_del-birnn4-scale5.0.dat};
\addlegendentry{BiRNN discount}
\addplot[very thick,color=red,mark=o] table [x=Dur,y=Del]{fig_used/swahili_evalpart1_manseg_dur_del.dat};
\addlegendentry{Minimum WER}
\end{axis}
\end{tikzpicture}
\caption{Impact of data selection schemes on deletion rate on narrow-band evaluation data}
\label{fig:eval_del}
\end{figure}
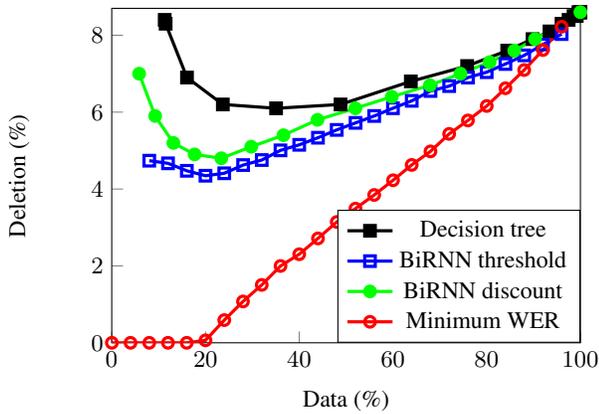

\begin{figure}[!htbp]
\centering
\begin{tikzpicture}
\begin{axis}[width=0.35\textwidth,height=0.25\textwidth,scale only axis,xlabel=Data (\%),ylabel=WER (\%),xmin=0,xmax=100,ymin=0.0,ymax=35.0,legend style={at={(1.0,0.0)},anchor=south east},xtick pos=left,ytick pos=left]
\addplot[very thick,color=black,mark=square*] table [x expr=100.0*\thisrow{D}/52293.55, y=WER]{fig_used/wer/swahili_evalpart1_manseg_map.dat};
\addlegendentry{Decision tree}
\addplot[very thick,color=blue,mark=square] table [x=Dur, y=WER]{fig_used/tmp.dat};
\addlegendentry{BiRNN threshold}
\addplot[very thick,color=green,mark=*] table [x expr=100.0*\thisrow{D}/52293.55, y=WER]{fig_used/wer/swahili_evalpart1_manseg_map_del-birnn4-scale5.0.dat};
\addlegendentry{BiRNN discount}
\addplot[very thick,color=red,mark=o] table [x=D, y=WER]{fig_used/swahili_dev_dur_wer.dat};
\addlegendentry{Minimum WER}
\end{axis}
\end{tikzpicture}
\caption{Impact of data selection approaches on total word error rate on narrow-band evaluation}
\label{fig:eval_wer}
\end{figure}
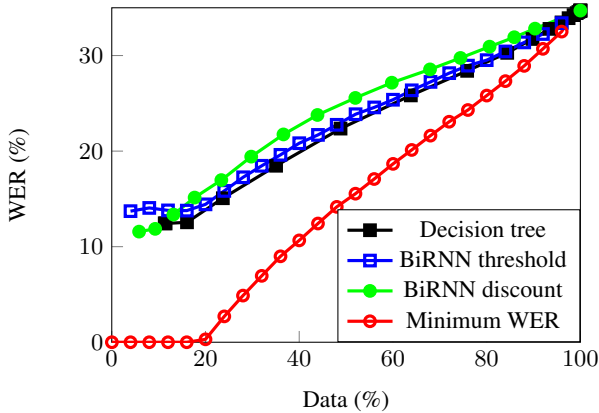

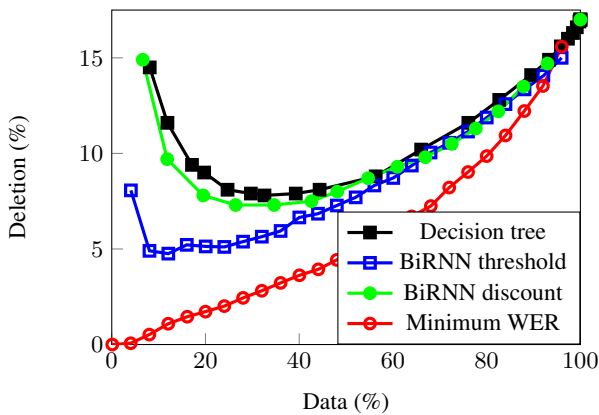
\begin{figure}[!htbp]
\centering
\begin{tikzpicture}
\begin{axis}[width=0.35\textwidth,height=0.25\textwidth,scale only axis,xlabel=Data (\%),ylabel=Deletion (\%),xmin=0,xmax=100,ymin=0.0,ymax=17.5,legend style={at={(1.0,0.0)},anchor=south east},xtick pos=left,ytick pos=left]
\addplot[very thick,color=black,mark=square*] table [x expr=100.0*\thisrow{D}/17515.08, y=Del]{fig_used/del/swahili_analysis1_map.dat};
\addlegendentry{Decision tree}
\addplot[very thick,color=blue,mark=square] table [x=Dur, y=Del]{fig_used/tmp3.dat};
\addlegendentry{BiRNN threshold}
\addplot[very thick,color=green,mark=*] table [x expr=100.0*\thisrow{Dur}/17515.08, y=Del]{fig_used/del/swahili_analysis1_wb_map_del-birnn4-scale5.0.dat};
\addlegendentry{BiRNN discount}
\addplot[very thick,color=red,mark=o] table [x=D, y=Del]{fig_used/del/1A_dur_wer_del.dat};
\addlegendentry{Minimum WER}
\end{axis}
\end{tikzpicture}
\caption{Impact of data selection schemes on deletion error rate on Material wide-band data}
\label{fig:analysis_del}
\end{figure}

The next experiment examined whether the AUC values in Table~\ref{tab:del_birnn} are high enough to enable untranscribed data selection with desirable error characteristics. Of particular interest in this work are transcriptions with low numbers of deletions to counter-balance the impact of domain mismatch between the narrow-band CTS Babel and wide-band news and topical broadcast Material data. For this investigation only Swahili was used. Two data selection schemes described in Section~\ref{sec:deletion} were investigated. The discounting coefficients in equations~\eqref{eq:discount1}--\eqref{eq:discount2} were estimated using a grid search. For simplicity a single discounting coefficient was estimated $\theta_d=\theta_s=5$. For the thresholding scheme in equations~\eqref{eq:threshold_cor}--\eqref{eq:threshold_del} the optimal thresholds were found by minimising mean squared error between the true WER and its estimate in equation~\eqref{eq:threshold_wer} using grid search. This yielded thresholds for correct/incorrect hypotheses $\theta_c=0.528$, deletion of the following word $\theta_d=0.941$ and deletion prior to the first hypothesised word $\theta_s=0.043$. Figure~\ref{fig:eval_del} shows how the number of deletions varies with the amount of data selected. Both schemes achieve lower deletion numbers for all quantities of the selected data with the thresholding scheme showing lowest numbers. Figure~\ref{fig:eval_wer} shows the impact these schemes have on WER.
The thresholding scheme largely follows the profile of the decision tree approach whereas the discounting scheme leads to higher WERs. As the parameters in these schemes were estimated on narrow-band data it is interesting to examine whether the estimated values would generalise to other domains. The equivalent of Figure~\ref{fig:eval_del} for the Material wide-band data is shown in Figure~\ref{fig:analysis_del}. This figure shows that the thresholding scheme generalised better to the wide-band data.

The final experiment examined whether any of these schemes could lead to WER improvements. As there is only general knowledge of the Material wide-band data domain composition (news and topical broadcast) a guess had to be made which of many possible Swahili wide-band audio files available on the internet to use. A subset of untranscribed Voice of America (Africa) data totalling 500 hours was selected and transcribed by the narrow-band recogniser from Table~\ref{tab:asr_cts}. A quarter of that data was chosen by each scheme to train a wide-band acoustic model. According to Figures~\ref{fig:eval_del} and \ref{fig:analysis_del} this should yield transcriptions with lower numbers of deletions. The WER performance of these models is compared in Table~\ref{tab:analysis_wer}. 
\begin{table}[!htbp]
\centering
\begin{tabular}{c||cccc}
\hline
\multirow{2}*{Selection} & \multicolumn{4}{c}{Error (\%)}\\
 & Sub & Del & Ins & Tot\\
\hline\hline
\text{narrow-band} & 23.4 & 19.0 & 1.3 & 43.7\\
\text{confidence} & 15.6 & 25.5 & 0.9 & 42.0\\
\text{threshold} & 16.5 & 25.2 & 1.1 & 42.8\\
\text{discount} & 16.9 & 23.7 & 1.1 & 41.7\\
\hline
\end{tabular}
\caption{Word error rate performance of data selection schemes on Material wide-band data}
\label{tab:analysis_wer}
\end{table}
The first line is the narrow-band recogniser. The next three lines shows the standard confidence-based scheme and BiRNN-based thresholding and discounting schemes. The standard confidence-based selection yields gains over the narrow-band recogniser although this comes with a significant increase in the number of deletion errors such that they become the dominant source of errors. The thresholding scheme shows worse generalisation from the VOA data compared to the standard confidence-based approach. This may be attributed to the mismatch between the VOA and Material data. The final line shows that the discounting scheme shows better generalisation. This may be attributed to a lesser degree of divergence between the curves of the decision tree and the discounting scheme in Figures~\ref{fig:eval_del}-\ref{fig:analysis_del} than it is for the thresholding scheme. Both schemes show better deletion error characteristics. Overall these results show promise in confidence and deletion-based data selection schemes and highlight challenges faced in building general purpose wide-band recognisers without any available transcribed data. 

\section{Conclusions}
\label{sec:conclusions}
This paper examined the problem of assigning accurate confidence scores to hypothesised words and predicting deletions produced by speech recognisers. The former, confidence estimation, problem has been researched before and a number of approaches have been proposed. This paper has examined bidirectional recurrent neural networks (BiRNN) - a powerful approach capable of incorporating information from both past and future words for predicting confidence in any given word. The latter, deletion prediction, problem has been researched less even though deletions are an important type of error that is hard to deal with in downstream and upstream tasks. This paper proposed a modification to the standard topology of BiRNNs to enable deletion prediction. The performance of BiRNNs was examined in challenging limited resource conditions across 3 different languages. The confidence prediction was found to yield better performance metrics than deletion prediction possibly due to the limited number of positive examples and features used in this exploration. The combination of the two predictions has been examined for unsupervised data selection in a highly mismatched domain. A simple discounting approach was found to provide small gains over the standard average frame-weighted confidence-based approach that does not take deletions into account. 

\bibliographystyle{IEEEbib}
\bibliography{references}

\end{document}